\documentclass[12pt]{iopart}
\usepackage{iopams}
\usepackage{graphicx}
\usepackage{epstopdf}
\usepackage{cite}
\usepackage{color}

\begin{document}

\title[Hole dynamics and spin currents]{Hole dynamics and spin currents after ionization in strong circularly polarized laser fields}
\author{Ingo Barth\footnote{Present address: Institut f\"ur Theoretische Physik and Centre for Quantum Engineering and Space-Time Research (QUEST), Leibniz Universit\"at Hannover, Appelstra\ss e 2, 30167 Hannover, Germany}
and Olga Smirnova}
\address{Max Born Institute, Max-Born-Str. 2A, 12489, Berlin, Germany}
\ead{ingo.barth@itp.uni-hannover.de, olga.smirnova@mbi-berlin.de}

\begin{abstract}
We apply the time-dependent analytical R-matrix theory to develop a movie of hole motion in a Kr atom upon ionization by strong circularly polarized field.
We find rich hole dynamics, ranging from rotation to swinging motion. The motion of the hole depends on the final energy and the spin  of the photoelectron and can be controlled  by  the  laser frequency and intensity. Crucially, hole rotation is a purely non-adiabatic effect, completely missing in the framework of quasistatic (adiabatic) tunneling theories.
We explore the possibility to use hole rotation as  a clock for measuring ionization time.
Analysing the relationship between the relative phases in different ionization channels we show that in the case of short-range electron-core interaction the hole is always initially aligned along the instantaneous direction of the laser field, signifying zero delays in ionization.
Finally, we show that strong-field ionization in circular fields  creates spin currents (i.e. different flow of spin-up and spin-down  density in space) in the ions. 
This phenomenon is intimately related to the  production of spin-polarized electrons in strong laser fields [Barth I and Smirnova O 2013 \textit{Phys. Rev. A} \textbf{88} 013401]. We demonstrate that rich spin dynamics of electrons and holes produced during strong field ionization can occur in typical experimental conditions and does not require relativistic intensities or strong magnetic fields. 
\end{abstract}
\submitto{\jpb}
\maketitle

\section{Introduction}

Time-resolving  attosecond photoionization dynamics  is at the forefront of current experimental \cite{ursi1,ursi2,Schultze,Krausz,ursi3,Klunder,Shafir,santr} and theoretical research 
 \cite{Dahlstr,Nagele,Thumm,Kheifets,Madsen,ivanov,suren,IvanovIA}.
 Experimental studies are enabled by tremendous progress in  modern technology, which allows one to remove an electron from an atom or a molecule quickly, and in a controlled way,
 using few-femtosecond intense infra-red (IR) pulses and/or attosecond extreme ultra violet (XUV) pulses \cite{Krausz2,Sanson}.
 Using these tools to develop a movie of non-equilibrium electron dynamics triggered by electron removal is one of the challenging goals of attosecond spectroscopy.

Non-equilibrium electron dynamics triggered by quick electron removal  corresponds to coherent population of different electronic states of the ion and can be followed by ultrafast hole migration \cite{hole1,hole2,hole3,hole4}.
 The dream of recording the movie of such motion  is now being brought closer to the reality by
the pioneering pump-probe  experiments in Kr \cite{Krausz,santr} atom and PENNA molecule \cite{milano} and the progress in high harmonic spectroscopy \cite{smirnova,pascal}.

The first frame of such movie should reveal the  initial shape and  momentum of the hole, in other words, the initial conditions of its motion.
 These initial conditions are set  by the process of electron rearrangement and are determined by the relative phases between the
  different ionization channels. In molecules, one could expect that longer, femtosecond time-scale nuclear dynamics that follows, may depend on these initial
   conditions due to an intricate interplay of electronic and nuclear degrees of freedom \cite{hole1,hole2,hole3,hole4,hole5}.


  Here we use theoretical tools \cite{Lisa1,Lisa2,jivesh,Lisa3} to develop the movie of hole motion in Kr atom upon ionization by strong circularly polarized fields.
  We focus
  on the problem of  ionization  phases between different ionization channels and their connection to time delays in ionization, including the prospect of using the hole motion as a clock for measuring ionization delays.
  We show that in the case of short-range electron-core interaction and two non-interacting ionization channels the hole is always initially aligned along the instantaneous direction of the laser field, signifying zero delays in ionization.
  
  We find that ionization of a noble gas atom by strong circularly polarized fields leads to rich hole dynamics, ranging from rotation to swinging motion.
  Crucially, the hole dynamics depends on the final energy and the spin  of the photoelectron and can be controlled  by changing the  laser frequency and intensity.
  Hole rotation  persists also after integration over the spin of the liberated electron. This effect owes its existence to the sensitivity of strong field ionization to the sense of electron rotation in the initial state, predicted in Refs. 
  \cite{PRA1,PRA2} and confirmed experimentally in Ref. \cite{Herath}.
  Theoretical \cite{PRA1,PRA2} and experimental \cite{Herath} results are in very good quantitative agreement \cite{comparison}.
   If the spin state of the liberated electron is not resolved, hole rotation is a purely non-adiabatic effect, vanishing as the Keldysh adiabaticity parameter $\gamma$ \cite{Keldysh} tends to zero.
  
  We show that strong field ionization creates different flow of spin-up and spin-down hole density, i.e. it creates spin current in the ion. 
  Up and down spin orientations are defined with respect to laser propagation direction. 
  The component of hole density with the spin oriented in the same direction as the spin of the removed electron encounters rich dynamics ranging from rotation to swinging motion. 
  The other component of the hole density with the spin opposite to the spin of the removed electron remains static, i.e its spatial shape does not evolve in time. Had strong-field ionization created equal number of spin-up and spin-down photoelectrons, there would be no spin current in the ion. However, ionization in strong circularly polarized field creates spin-polarized electrons \cite{PRAspin}. Thus, it will also create preferential direction of the spin current in the ion, 
  even after integration over the spin of the liberated electron.

   Spin dynamics in relativistic ionization has been recently considered by Klaiber et al \cite{Klaiber}. Ref \cite{Klaiber} shows that the magnetic-field component of the linearly polarized super-strong laser field with intensity larger than $10^{20}$\,Wcm$^{-2}$ can flip the electron spin in hydrogen-like systems.
  Combining laser fields with strong magnetic fields, Refs.\,\cite{Dirac1,Dirac2} explore  relativistic spin currents in cycloatoms.
  
   We demonstrate that rich spin dynamics can occur in  typical experimental conditions and does not require relativistic intensities or strong magnetic fields.
  
  To derive the analytical expressions for the time-dependent hole densities, we depart from the frequency-domain approach used by us in Refs.\,\cite{PRA1,PRA2,PRAspin}, similar to the PPT theory \cite{PPT1}, and use the results of time-dependent analytical R-matrix method (ARM),  Ref.\,\cite{Lisa1,Lisa2,jivesh,Lisa3}.
  Pertinent theoretical work, focussing on numerical approaches to multichannel dynamics of strong field ionization includes Refs.\,\cite{SantraKr,nina,michael}.


  The paper is organized as follows. In \sref{ii} we derive expressions for channel-specific ionization amplitudes.
   In \sref{iii} we derive analytical expression for the hole density depending on the spin and  the energy of the photoelectron. 
   In \sref{iv} we discuss  the initial alignment of the hole and the possibility to use it for the detection of the electron emission times.
   In \sref{v} we show that the dynamics of spin-up and spin-down components of the hole is different leading to spin currents in the ion.
  In \sref{vi} we develop the movie of hole motion for specific parameters of the circularly polarized laser field. 
  Section \sref{vii} concludes the work.

\section{Ionization amplitudes}
\label{ii}
We shall consider hole formation after a single ionization event, i.e. the hole formed by an ionization burst during one cycle of the infrared field.
The ionization amplitude for the spin-less electron in general case of an arbitrary binding potential has been derived in Ref.\,\cite{jivesh} (equation (67)) using time-dependent analytical R-matrix method (ARM).
Here we generalize this expression to take into account the electron spin.

To get insight into the electron or hole dynamics in a noble gas  ion, such as Kr, after ionization by right ($c=+1$) or left ($c=-1$) circularly polarized laser fields,
we start with the total time-dependent wavefunction of the ion and of the photoelectron including spin $|\Psi_c(t)\rangle$ \cite{Lisa2}.
The wavefunction of the remaining ion, where the photoelectron has final momentum $\mathbf{p}$ and spin $m_s=\pm\frac12$, is obtained by
projecting the total spatial and spin wavefunction of the photoelectron $\langle\mathbf{p}\chi_{\frac12m_s}|$ onto the total wavefunction $|\Psi_c(t)\rangle$ describing the ionization, i.e.
\begin{eqnarray}
\label{Psi1}
|\Psi_c^{m_s}(\mathbf{p},t)\rangle&=&\langle \mathbf{p}\chi_{\frac12m_s}|\Psi_c(t)\rangle.
\end{eqnarray}
We have to consider six ionization channels, corresponding to leaving the ion in its six lowest eigenstates  $|^2\!P_{JM_J}\rangle$,
\begin{eqnarray}
\label{Psi2}
|\Psi_c^{m_s}(\mathbf{p},t)\rangle&=&\sum_{J,M_J}a_c^{JM_Jm_s}(\mathbf{p},t)|^2\!P_{JM_J}\rangle,
\end{eqnarray}
where
\begin{eqnarray}
\label{aJMJ1}
a_c^{JM_Jm_s}(\mathbf{p},t)&=&\langle ^2\!P_{JM_J}\mathbf{p}\chi_{\frac12m_s}|\Psi_c(t)\rangle
\end{eqnarray}
are the channel-specific ionization amplitudes.
We note that the quantum numbers in uppercase and in lowercase  correspond to the quantum numbers assigned to the ion and to the photoelectron, respectively.
Within the ARM formalism, the total wavefunction $|\Psi_c(t)\rangle$ 
 in the outer region of the 
R-matrix sphere is given by \cite{Lisa2}
\begin{eqnarray}
\label{Psict}
|\Psi_c(t)\rangle&=&-i\sum_{J,M_J,m'_s}\int d\mathbf{k}\int_{-\infty}^t dt'\,U(t,t')|^2\!P_{JM_J}\mathbf{k}\chi_{\frac12m'_s}\rangle\\\nonumber
&&\qquad\qquad\qquad\qquad\langle ^2\!P_{JM_J}\mathbf{k}\chi_{\frac12m'_s}|\hat\Delta(a)\hat B|^1\!S_0\rangle a_0(t')e^{-iE_0t'} .
\end{eqnarray}
Here $U(t,t')$ is the full propagator for the atom in the laser field,
$\hat\Delta(a)\hat B$ is the Bloch operator (for definition and details, including explicit expressions for  $\hat\Delta(a)\hat B$  see Ref.\,\cite{Lisa2}),
$|^1\!S_0\rangle$ is the  ground state of the neutral noble gas atom prior to ionization, $a_0(t')$ incorporates Stark shift and the depletion of the ground state, and $E_0$ is the ground state energy. Contribution of the exchange terms in the outer R-matrix region into the optical tunneling amplitude are small (Ref.\,\cite{Lisa2}) and are neglected here.
 We approximate the full propagator $U(t,t')$  as $U(t,t')=U^{ion}(t,t')U^e(t,t')$,  where
$U^{ion}(t,t')$ and $U^e(t,t')$ are the propagators for the ion and for the photoelectron (see for details Ref.\,\cite{Lisa2}), respectively, thus neglecting the correlation-driven processes, such as those considered in Ref.\,\cite{Lisa2}.
The propagator for the ion uses exact ionic eigenstates which
account for the spin-orbit interaction in the ion, see
below. For the continuum electron, the spin-orbit interaction
in the outer region is neglected.

Since the laser field does not couple the lowest ionic states $|^2\!P_{JM_J}\rangle$ of a noble gas ion directly, one would expect to observe similar Stark shifts for these states. Calculated   in Ref.\,\cite{SantraKr} for  the linearly polarized field of the strength $\mathcal{E}$=0.1 a.u., these Stark shifts lead to the overall change in the energy spacing between $|^2\!P_{\frac12M_J}\rangle$ and  $|^2\!P_{\frac32M_J}\rangle$ at the level of $7\%$. 
In this work we consider  weaker fields $\mathcal{E}\le0.05$\,a.u., and therefore we expect negligible modifications of the energy spacing due to the Stark shifts.
Neglecting the relative Stark shifts of the ionic states, we obtain $\langle ^2\!P_{JM_J}|U^{ion}(t,t')=\langle ^2\!P_{JM_J}|e^{-iE_J(t-t')}$,
where $E_J$ are the  eigenenergies of the ionic states $|^2\!P_{JM_J}\rangle$. Since the propagator $U^e(t,t')$ and the Bloch operator $\hat\Delta(a)\hat B$ do not depend on the spin of the photoelectron, we can rewrite
equation \eref{aJMJ1} as
\begin{eqnarray}
\label{aJMJ2}
a_c^{JM_Jm_s}(\mathbf{p},t)&=&-i\int d\mathbf{k}\int_{-\infty}^t dt'\,\langle\mathbf{p}|U^e(t,t')|\mathbf{k}\rangle\\\nonumber
&&\qquad\langle ^2\!P_{JM_J}\mathbf{k}\chi_{\frac12m_s}|\hat\Delta(a)\hat B|^1\!S_0\rangle a_0(t')e^{-iE_0t'}e^{-iE_J(t-t')}.
\end{eqnarray}
In the  Koopmans' approximation, the Dyson orbital $\langle ^2\!P_{JM_J}|^1\!S_0\rangle=|p_{J-M_J}\rangle$, 
where $|p_{J-M_J}\rangle$ is the valence spin-orbital of the noble gas atom,
because $|^2\!P_{JM_J}\rangle$ does not contain the spin-orbital $p_{J-M_J}$, whereas $|^1\!S_0\rangle$ contains all $s$ and $p$ spin-orbitals.
Using the Clebsch-Gordan expansion for spin orbitals $p_{JM_J}$ in the basis of the products of spatial orbitals $p_M=p_{M_J-M_S}$ and the spin functions $\chi_{\frac12 M_s}$, i.e.
\begin{eqnarray}
p_{JM_J}(\mathbf{r},\sigma)&=&\sum_{M_S} C_{1\,M_J-M_S,\frac12 M_S}^{JM_J}p_{M_J-M_S}(\mathbf{r})\chi_{\frac12 M_S}(\sigma),
\label{pJMJ}
\end{eqnarray}
the formula for the ionization amplitudes \eref{aJMJ2} can be rewritten as
\begin{eqnarray}
\label{aJMJ3}
a_c^{JM_Jm_s}(\mathbf{p},t)&=&-iC_{1\,-M_J-m_s,\frac12\,m_s}^{J-M_J}\int d\mathbf{k}\int_{-\infty}^t dt'\,\langle\mathbf{p}|U^e(t,t')|\mathbf{k}\rangle\\\nonumber
&&\qquad\langle \mathbf{k}|\hat\Delta(a)\hat B|p_{-M_J-m_s}\rangle a_0(t')e^{-iE_0t'}e^{-iE_J(t-t')}.
\end{eqnarray}
Using the definition for the ionization amplitudes for the atomic orbitals (cf.\ equation \eref{Psict})
\begin{eqnarray}\nonumber
a_c^{p_m}(\mathbf{p},t,I_p^{P_J})&=&-i\int d\mathbf{k}\int_{-\infty}^t dt'\,\langle\mathbf{p}|U^e(t,t')|\mathbf{k}\rangle\langle\mathbf{k}|\hat\Delta(a)\hat B|p_m\rangle a_0(t')e^{iI_p^{P_J}t'}\\
\label{acpm}
\end{eqnarray}
with the ionization potential $I_p^{P_J}=E_J-E_0$, we obtain the final formula for the ionization amplitudes
\begin{eqnarray}
\label{afinal}
a_c^{JM_Jm_s}(\mathbf{p},t)&=&C_{1\,-M_J-m_s,\frac12\,m_s}^{J-M_J}a_c^{p_{-M_J-m_s}}(\mathbf{p},t,I_p^{P_J}) e^{-iE_Jt}.
\end{eqnarray}

%

\section{Analytical expressions for the hole density}
\label{iii}
The hole density can be defined \cite{nina} as:
\begin{eqnarray}
\label{holedensity}
\rho_c^{m_s}(\mathbf{r},\mathbf{p},\sigma,t)&=&\sum_{J,M_J,J',M_J'}a_c^{JM_Jm_s}(\mathbf{p},t)\left[p_{J-M_J}(\mathbf{r},\sigma)\right]^{*}\\\nonumber
&&\qquad\qquad\left[a_c^{J'M_J'm_s}(\mathbf{p},t)\right]^{*} p_{J'-M_J'}(\mathbf{r},\sigma),
\end{eqnarray}
where   $a_c^{JM_Jm_s}(\mathbf{p},t)$ are the population amplitudes of the ionic states $|^2\!P_{JM_J}\rangle$ (see equation \eref{afinal}),
$p_{J-M_J}(\mathbf{r},\sigma)$ are the  valence spin-orbitals, $\sigma$ characterizes  spin variables of the hole and $m_s$ describes the spin of the photoelectron.
To obtain simple expressions for  hole density \eref{holedensity},
we can neglect very small ionization rates or amplitudes for $p_0$ orbitals, cf.\ Refs.\cite{PRA1,PRA2}, i.e.
\begin{eqnarray}
\label{a0}
a_c^{p_0}(\mathbf{p},t,I_p)&\approx&0.
\end{eqnarray}
Using equations \eref{pJMJ} and \eref{afinal}, one can obtain the components of the hole density (see \tref{aproducttab1} for the spin-up electron and \tref{aproducttab2} for the spin-down electron),  where we use the abbreviation
$ a_c^{p_m}(\mathbf{p},t,I_p^{P_J})=a_c^{p_mJ}$. 

\begin{table}
\caption{\label{aproducttab1}Elements of hole density (equation \eref{holedensity}) for the spin-up electron. The elements for $J',M_J'$ are the same but conjugated.}
\begin{indented}
\renewcommand{\arraystretch}{2}
\item[]\begin{tabular}{@{}ccc}
\br
$(J,M_J)$ & $\left[p_{J-M_J}(\mathbf{r},\sigma)\right]^{*}$ & $a_c^{JM_Jm_s}(\mathbf{p},t)$  \\
\mr
$(\frac12,\frac12)$ & $\left[\frac{1}{\sqrt{3}}p_0^{\downarrow}-\sqrt{\frac{2}{{3}}}p_-^{\uparrow} \right]^{*}$ & $-\sqrt{\frac{2}{{3}}}a_c^{p_{-}\frac12} e^{-iE_{\frac12}t}$  \\\hline
$(\frac32,\frac12)$ & $\left[\sqrt{\frac{2}{{3}}}p_0^{\downarrow}+\frac{1}{\sqrt{3}}p_-^{\uparrow} \right]^{*}$  & $\frac{1}{\sqrt{3}}a_c^{p_{-}\frac32} e^{-iE_{\frac32}t}$  \\\hline
$(\frac32,-\frac32)$ & $\left[p_{+}^{\uparrow}\right]^{*}$ & $a_c^{p_{+}\frac32} e^{-iE_{\frac32}t}$\\
\br
\end{tabular}
\end{indented}
\end{table}
\begin{table}
\caption{\label{aproducttab2}Elements of hole density (equation \eref{holedensity}) for the spin-down electron. The elements for $J',M_J'$ are the same but conjugated.}
\begin{indented}
\renewcommand{\arraystretch}{2.5}
\item[]\begin{tabular}{ccc}
\br
$(J,M_J)$ & $\left[p_{J-M_J}(\mathbf{r},\sigma)\right]^{*}$ & $a_c^{JM_Jm_s}(\mathbf{p},t)$  \\
\mr
$(\frac12,-\frac12)$ & $\left[-\frac{1}{\sqrt{3}}p_0^{\uparrow}+\sqrt{\frac{2}{{3}}}p_+^{\downarrow} \right]^{*}$ & $\sqrt{\frac{2}{{3}}}a_c^{p_{+}\frac12} e^{-iE_{\frac12}t}$  \\\hline
$(\frac32,-\frac12)$ & $\left[\sqrt{\frac{2}{{3}}}p_0^{\uparrow}+\frac{1}{\sqrt{3}}p_+^{\downarrow} \right]^{*}$  & $\frac{1}{\sqrt{3}}a_c^{p_{+}\frac32} e^{-iE_{\frac32}t}$  \\\hline
$(\frac32,\frac32)$ & $\left[p_{-}^{\downarrow}\right]^{*}$ & $a_c^{p_{-}\frac32} e^{-iE_{\frac32}t}$\\
\br
\end{tabular}
\end{indented}
\end{table}
Since the elements of the hole density (equation \eref{holedensity}) shown in \tref{aproducttab1} and \tref{aproducttab2} are complex, the hole dynamics crucially depends on the relative phases between these elements. The phases of the spatial parts are defined as follows:  
$p_\pm(\mathbf{r})=\mp|p_\pm(\mathbf{r})|e^{\pm i\phi}$ \cite{footnote}, where $\phi$ is the azimuthal angle. 

The following products of ionization amplitudes contribute to the hole density: 
\begin{eqnarray}
\left(a_c^{p_{m'}}(\mathbf{p},t,I_p^{P_{J'}})\right)^*a_c^{p_m}(\mathbf{p},t,I_p^{P_J})
&=&\left|a_c^{p_{m'}}(\mathbf{p},t,I_p^{P_{J'}})\right|\left|a_c^{p_m}(\mathbf{p},t,I_p^{P_J})\right|\\\nonumber
&&e^{i\left(\eta_c^{p_m}(\mathbf{p},t,I_p^{P_J})-\eta_c^{p_{m'}}(\mathbf{p},t,I_p^{P_{J'}})\right)},
\end{eqnarray}
where $\eta_c^{p_m}(\mathbf{p},t,I_p^{P_J})$ and $\eta_c^{p_{m'}}(\mathbf{p},t,I_p^{P_{J'}})$ are the corresponding phases of the ionization amplitudes.
As expected,  only the relative phases between different ionization channels contribute to the final expressions.
\begin{eqnarray}
\label{etarel}
\Delta\eta_c^{p_mp_{m'}}(\mathbf{p},t,I_p^{P_J},I_p^{P_{J'}})&=&\eta_c^{p_m}(\mathbf{p},t,I_p^{P_J})-\eta_c^{p_{m'}}(\mathbf{p},t,I_p^{P_{J'}}).
\end{eqnarray}
Following the standard routine of the ARM approach \cite{jivesh,Lisa3}, we apply the saddle point method to equation \eref{acpm} and obtain the analytical expression for the ionization amplitude $a_c^{p_m}(\mathbf{p},t,I_p^{P_J})$. 
The result is proportional to $e^{iI_p^{P_J}t_i}$, where $t_i$ is the complex ionization time (saddle point of the time integral in equation \eref{acpm}). The real component of this time $t_0=\mathrm{Re}\,t_i$ is known as the time when the electron is liberated.
Since the ionization potential appears only in this factor and this factor does not depend on $m$, the relative phase \eref{etarel} is then approximately expressed as
\begin{eqnarray}
\nonumber
 \Delta\eta_c^{p_mp_{m'}}(\mathbf{p},t,I_p^{P_J},I_p^{P_{J'}})&\approx& \Delta\eta(t_i,I_p^{P_J},I_p^{P_{J'}})+\Delta\eta^{p_mp_{m'}}(\mathbf{p},t_i)=\Delta\eta_1+\Delta\eta_2.\\
 \label{etaapprox}
\end{eqnarray}
The first term of equation \eref{etaapprox} is evaluated as
\begin{eqnarray}
\label{eta1}
\Delta\eta_1=\Delta\eta(t_i,I_p^{P_J},I_p^{P_{J'}})&=&I_p^{P_J}\,\mathrm{Re}\,t_i-I_p^{P_{J'}}\,\mathrm{Re}\,t_i=E_{SO}^{JJ'}\,\mathrm{Re}\,t_i,
\end{eqnarray}
where $E_{SO}^{JJ'}=-E_{SO}^{J'J}=I_p^{P_J}-I_p^{P_{J'}}=E_J-E_{J'}$ is the spin-orbit energy splitting. To simplify the notations, we denote $E_{SO}^{\frac12\frac32}=E_{SO}$.
We prove in the Appendix that the second term of equation \eref{etaapprox} for $m,m'=\pm1$ is
\begin{eqnarray}
\label{eta2}
\Delta\eta_2=\Delta\eta^{p_mp_{m'}}(\mathbf{p},t_i)&=&(m-m')\theta_p,
\end{eqnarray}
where $\theta_p$ is the photoelectron detection angle or equivalently the azimuthal angle of the electron final momentum $\mathbf{p}$.
An alternative derivation of the ionization phase, which additionally includes the Coulomb effects can be found in \cite{jivesh}. The results of these two derivations agree for short-range potentials.
It is interesting to note that the relative phases $\Delta\eta_1$ and $\Delta\eta_2$ do not depend on the sense of circular polarization.

The relative phase $\Delta\eta_1$ is accumulated due to the energy splitting between the two lowest electronic states of the ion.
The relative phase $\Delta\eta_2$ is accumulated between different ionization channels corresponding to the removal of co-rotating and counter-rotating electron correspondingly.

Thus, our final result is expressed via ionization amplitudes for $p_{+}$ and $p_{-}$ electrons, calculated for ionization potentials corresponding to the two lowest states of the ion with $J=\frac32$ and $J=\frac12$.
The absolute values of these ionization amplitudes $\left|a_c^{p_mJ}\right|$ have been derived in Refs.\,\cite{PRA1,PRA2}.

With equations \eref{etarel}--\eref{eta2}, $p_\pm(\mathbf{r})=\mp|p_\pm(\mathbf{r})|e^{\pm i\phi}$, \tref{aproducttab1}, and \tref{aproducttab2}, we obtain the final analytical expression for the hole density \eref{holedensity}
depending on the spin of the photoelectron $m_s=\pm\frac12$, i.e.
\begin{eqnarray}
\label{holedensityfinal}
\Delta\rho_c^{\pm\frac12}(\mathbf{r},\mathbf{p},t)&\approx&
\frac29\left|a_c^{p_\mp\frac12}\right|^2\left(|p_0|^2+2|p_\mp|^2\right)\\\nonumber
&&+\frac19\left|a_c^{p_\mp\frac32}\right|^2\left(2|p_0|^2+|p_\mp|^2\right)+\left|a_c^{p_\pm\frac32}\right|^2|p_\pm|^2\\\nonumber
&&-\frac49\left|a_c^{p_\mp\frac12}\right|\left|a_c^{p_\mp\frac32}\right|\left(|p_0|^2-|p_\mp|^2\right)\cos\left(E_{SO}\left(t-\mathrm{Re}\,t_i\right)\right)\\\nonumber
&&-\frac23\left|a_c^{p_\mp\frac32}\right|\left|a_c^{p_\pm\frac32}\right||p_\pm|^2\cos\left(2\left(\phi-\theta_p\right)\right)\\\nonumber
&&-\frac43\left|a_c^{p_\mp\frac12}\right|\left|a_c^{p_\pm\frac32}\right||p_\pm|^2\cos\left(E_{SO}\left(t-\mathrm{Re}\,t_i\right)\mp2\left(\phi-\theta_p\right)\right).
\end{eqnarray}
The last term in equation \eref{holedensityfinal} describes the rotation of the hole, with the initial phase $2\left(\phi-\theta_p\right)$.
This part of the interchannel phase reflects the correlation between the initial alignment of the hole and the angle of electron detection.


To complete the analysis of equation \eref{holedensityfinal}, we note the following symmetry rules.
 Naturally, the magnitudes of the ionization amplitudes for $p_{-}$ electron in right circularly polarized laser fields $|a_+^{p_-J}|$ is equal to the one for $p_{+}$ electron in left circularly polarized field $|a_-^{p_+J}|$ (cf.\,Refs.\,\cite{PRA1,PRA2}). It is also natural that the hole dynamics for right and left circularly polarized laser fields are not identical, i.e.
$\Delta\rho_+^{\pm\frac12}(\mathbf{r},\mathbf{p},t)\neq\Delta\rho_-^{\mp\frac12}(\mathbf{r},\mathbf{p},t)$, but correspond to the opposite sense of rotation for the hole as can be easily seen from
 equation \eref{holedensityfinal}.

\section{Initial alignment of the hole}
\label{iv}
Equation \eref{holedensityfinal} shows that the hole created upon ionization rotates, i.e.
the maximum of the hole density points at different directions at different times.
How is the hole aligned with respect to the laser field when it is just created?
Does the initial hole alignment follow the rotation of the laser field, i.e. does the attoclock principle \cite{ursi1,ursi2,ursi3} also apply to the hole?
Can one detect ionization time by probing the initial shape of the hole?

The short-range interaction between the photoelectron and the core considered here is the perfect testbed to compare both clocks:
 the attoclock operating on the electron and the internal clock operating on the rotating hole.
The attoclock principle maps the angle of electron detection onto the ionization time. The mapping is simple and unambiguous for a short range potential.
In this case, the electron is ejected along the instantaneous direction of the laser field. The time of electron emission $t_0$ is mapped onto the electron detection angle $\theta_p$ as $\omega t_0=\frac{\pi}{2}+\mathrm{sgn}(c)\,\theta_p$.
Is there a similar connection between the angle $\phi$ corresponding to the maximum of the hole density and the ionization time?
Intuitively, one would expect that the hole density is maximum at the same angle at which the electron has been removed, i.e. along the instantaneous direction of the laser field.
Initial mismatch  between these directions  could indicate ionization delays.
Since ionization delay is strictly zero for short range potentials \cite{archive}, we expect the hole to be aligned along the direction of the laser field.

Let us use equation \eref{holedensityfinal} to look at the initial shape of the hole created at the time $t=\mathrm{Re}\,t_i\equiv t_0$, when according to the short-range theory (see e.g. Refs.\,\cite{PRA1,PRA2}) the electron exits the barrier and becomes free:
\begin{eqnarray}
\label{holedensityinitial}
\Delta\rho_c^{\pm\frac12}(\mathbf{r},\mathbf{p},\mathrm{Re}\,t_i)&\approx&
\frac29\left|a_c^{p_\mp\frac12}\right|^2\left(|p_0|^2+2|p_\pm|^2\right)\\\nonumber
&&+\frac19\left|a_c^{p_\mp\frac32}\right|^2\left(2|p_0|^2+|p_\pm|^2\right)+\left|a_c^{p_\pm\frac32}\right|^2|p_\pm|^2\\\nonumber
&&-\frac49\left|a_c^{p_\mp\frac12}\right|\left|a_c^{p_\mp\frac32}\right|\left(|p_0|^2-|p_\pm|^2\right)\\\nonumber
&&-\frac23\left(2\left|a_c^{p_\mp\frac12}\right|+\left|a_c^{p_\mp\frac32}\right|\right)\left|a_c^{p_\pm\frac32}\right||p_\pm|^2\cos\left(2\left(\phi-\theta_p\right)\right).
\end{eqnarray}

The initial alignment of the hole is determined by the last term of equation \eref{holedensityinitial}.
Since the sign of this term is negative, the hole density at the initial time $t=\mathrm{Re}\,t_i=\frac{\pi}{2}+\mathrm{sgn}(c)\,\theta_p$  is maximal at $\phi=\pm\frac{\pi}{2}+\theta_p$, i.e.\ the hole is aligned with the laser field, in agreement with the argument above.
Indeed, according to the short-range theory, the photoelectron leaves the barrier at $\phi_e=\theta_p-\mathrm{sgn}(c)\,\frac{\pi}{2}$ and arrives at the detector at the  angle $\theta_p$, see figure 1 of Ref.\,\cite{PRA1}.

Thus, we conclude, that in the case of short-range potential considered here, the initial alignment of the hole can  be directly linked to the absence of time delays in its formation. Note that this result can not be directly ported to long-range potentials.

We note in passing, that in contrast to the hole dynamics \eref{holedensityfinal}, the initial hole for right circular polarization and $m_s=\pm\frac12$ has the same shape as the one for left circular polarization and $m_s=\mp\frac12$,
i.e.\,$\Delta\rho_+^{\pm\frac12}(\mathbf{r},\mathbf{p},\mathrm{Re}\,t_i)=\Delta\rho_-^{\mp\frac12}(\mathbf{r},\mathbf{p},\mathrm{Re}\,t_i)$.

\section{Spin currents}
 \label{v}
 Instead of integrating over the hole spin, one could consider the dynamics of spin-up and spin-down hole density separately. 
 Consider  the case of the spin-up electron.
 The spin-down component of the hole density does not evolve in space:
 \begin{eqnarray}
 \label{holedensityfinal_down}
\Delta\rho_c^{\frac12,\downarrow}(\mathbf{r},\mathbf{p},t)&\approx&
\frac29\left|a_c^{p_-\frac12}\right|^2|p_0|^2\\\nonumber
&&
\left[1+\frac{\left|a_c^{p_-\frac32}\right|^2}{\left|a_c^{p_-\frac12}\right|^2}
-2\frac{\left|a_c^{p_-\frac32}\right|}{\left|a_c^{p_-\frac12}\right|}\cos\left(E_{SO}\left(t-\mathrm{Re}\,t_i\right)\right)\right],
\end{eqnarray}
 while the spin-up component of the hole density,
 \begin{eqnarray}
 \label{holedensityfinal_up}
\Delta\rho_c^{\frac12,\uparrow}(\mathbf{r},\mathbf{p},t)&\approx&
\frac49\left|a_c^{p_-\frac12}\right|^2|p_-|^2+\frac{1}{9}\left|a_c^{p_-\frac32}\right|^2|p_-|^2+\left|a_c^{p_+\frac32}\right|^2|p_+|^2\\\nonumber
&&+\frac49\left|a_c^{p_-\frac12}\right|\left|a_c^{p_-\frac32}\right||p_-|^2\cos\left(E_{SO}\left(t-\mathrm{Re}\,t_i\right)\right)\\\nonumber
&&-\frac23\left|a_c^{p_-\frac32}\right|\left|a_c^{p_+\frac32}\right||p_+|^2\cos\left(2\left(\phi-\theta_p\right)\right)\\\nonumber
&&-\frac43\left|a_c^{p_-\frac12}\right|\left|a_c^{p_+\frac32}\right||p_+|^2\cos\left(E_{SO}\left(t-\mathrm{Re}\,t_i\right)-2\left(\phi-\theta_p\right)\right),
\end{eqnarray}
rotates due to the presence of the last term in the equation \eref{holedensityfinal_up}.

We can make two conclusions. First,  strong field ionization creates different flow of spin-up and spin-down hole density, i.e.\ 
creates spin currents in the ion ( calculation of the magnitude of spin currents is outside the scope of this work). 
Second, the spatially rotating component of the hole density correlates to the spin of the removed electron 
(i.e.\ the spin-up component of the hole density for the spin-up electron). 
Since ionization in strong circularly polarized field creates spin-polarized electrons \cite{PRAspin}, it will also create preferential direction of the spin current in the ion even after integration over the electron spin. 

 Eqs.\eref{holedensityfinal_down},\eref{holedensityfinal_up} can be understood within the following intuitive physical picture.
Removal of an electron with spin up should initially create a hole in the  spin up orbital (the lack of the spin-up electron). However, the appearance of  the hole in the spin down orbital in Eq.\eref{holedensityfinal_down} signifies  the 
spin-flip process caused by the spin-orbit interaction. Indeed, the component of the  hole density in the spin down orbital vanishes at the initial moment of time $\mathrm{Re}\,t_i$  (and remains equal to zero at all times) if we set $E_{SO}=0$, see  Eq. \eref{holedensityfinal_down}). The spin-flip must conserve the total angular momentum. Therefore, reducing $m_s$  by one (from $\frac{1}{2}$ to $-\frac{1}{2}$) requires the increase of $m_l$ by one, which is only possible in case of  the hole initially in the $p_-$ orbital. Due to increase of $m_l$ by one the initial $p_-$ hole turns into $p_0$ hole.  Thus, Eq. \eref{holedensityfinal_down} shows that the  hole in the spin down orbital can only contain $p_0$ orbital, with a time-dependent weight determined exclusively by the ionization probability corresponding to electron removal from $p_-$ orbital. Holes produced by removal of spin up $p_+$ electrons can not be involved in the spin-flip due to momentum conservation and therefore can not contribute to the spin-down component of the hole density given by Eq.\eref{holedensityfinal_down}. Thus, there is no spatial dynamics  in the hole density  in the spin-down orbital, only its weight is time-dependent due to the spin-orbit dynamics.
In contrast,  the hole in the spin up orbital in  Eq. \eref{holedensityfinal_up} contains both $p_+$ and $p_-$ orbitals, which carry opposite currents, enabling the transport of density in space.

In case of spin-down electron removal one obtains similar expressions for spin-resolved hole density. The spin-up component of the hole density does not rotate:
 \begin{eqnarray}
\Delta\rho_c^{-\frac12,\uparrow}(\mathbf{r},\mathbf{p},t)&\approx&
\frac29\left|a_c^{p_+\frac12}\right|^2|p_0|^2\\\nonumber
&&\left[1+\frac{\left|a_c^{p_+\frac32}\right|^2}{\left|a_c^{p_+\frac12}\right|^2}
-2\frac{\left|a_c^{p_+\frac32}\right|}{\left|a_c^{p_+\frac12}\right|}\cos\left(E_{SO}\left(t-\mathrm{Re}\,t_i\right)\right)\right],
\label{dholedensityfinal_down1}
\end{eqnarray}
 The  spin-down component of the hole density rotates in space:
 \begin{eqnarray}
 \label{dholedensityfinal_up1}
\Delta\rho_c^{-\frac12,\downarrow}(\mathbf{r},\mathbf{p},t)&\approx&
\frac49\left|a_c^{p_+\frac12}\right|^2|p_+|^2+\frac{1}{9}\left|a_c^{p_+\frac32}\right|^2|p_+|^2+\left|a_c^{p_-\frac32}\right|^2|p_-|^2\\\nonumber
&&+\frac49\left|a_c^{p_+\frac12}\right|\left|a_c^{p_+\frac32}\right||p_+|^2\cos\left(E_{SO}\left(t-\mathrm{Re}\,t_i\right)\right)\\\nonumber
&&-\frac23\left|a_c^{p_+\frac32}\right|\left|a_c^{p_-\frac32}\right||p_-|^2\cos\left(2\left(\phi-\theta_p\right)\right)\\\nonumber
&&-\frac43\left|a_c^{p_+\frac12}\right|\left|a_c^{p_-\frac32}\right||p_-|^2\cos\left(E_{SO}\left(t-\mathrm{Re}\,t_i\right)\!+\!2\left(\phi-\theta_p\right)\right).
\end{eqnarray}

\section{Movie of the hole dynamics}
\label{vi}
In this section we
illustrate our results using the example of a krypton atom subject to strong circularly polarized laser field.
The ionization channel from the ground neutral state $|^1\!S_0\rangle$ to the final ionic state $|^2\!P_{\frac32M_J}\rangle$ or $|^2\!P_{\frac12M_J}\rangle$
corresponds to the removal of an electron from the valence spin-orbital $4p_{\frac32m_j}$ or $4p_{\frac12m_j}$ with $j=J$ and $m_j=-M_J$.
The corresponding ionization potentials are  $I_p^{P_{3/2}}=0.5145\,$a.u.\ and $I_p^{P_{1/2}}=0.5389\,$a.u.\,\cite{NIST},
corresponding to the spin-orbit splitting $E_{SO}=0.0245$\,a.u.\,$=0.67$\,eV with period of 6.2\,fs.


We consider the hole dynamics resolved on the final energy and spin of the ejected electron.
Due to right/left symmetry, we only consider a right circularly polarized laser field ($c=+1$) and use two different laser amplitudes $\mathcal{E}=0.05\,$a.u.\ for 800\,nm and $\mathcal{E}=0.02\,$a.u.\ for 1600\,nm.
We select the final kinetic energy of the photoelectron $E_\mathrm{kin}=0.70\,$a.u.\ for 800\,nm and $E_\mathrm{kin}=0.45\,$a.u.\ for 1600\,nm.
Two different spins of the photoelectron  $m_s=\pm\frac12$ lead to different scenarios of the movies for the spin-resolved
hole dynamics in the ion (equation \eref{holedensityfinal}) with corresponding initial frames (equation \eref{holedensityinitial}),
that also depend on the laser frequency, the laser amplitude, and the final electronic kinetic energy.
Note, that the hole dynamics \eref{holedensityfinal} is determined by the magnitudes of the ionization amplitudes $|a_c^{p_\mp\frac12}|$, $|a_c^{p_\mp\frac32}|$, $|a_c^{p_\pm\frac32}|$.
Since the norm of the hole density is irrelevant for its dynamics, we can divide equation \eref{holedensityfinal} by $|a_c^{p_\mp\frac32}|^2$.
It yields the expression for the hole density that depends only on the two ratios of the ionization amplitudes or, approximately, on the two corresponding ratios for the roots of the energy-resolved ionization rates, i.e.
\begin{eqnarray}
\label{a1}
\frac{|a_c^{p_\mp\frac12}|}{|a_c^{p_\mp\frac32}|}&\approx&\sqrt{\frac{w_{nc}^{p_\mp}(\mathcal{E},\omega,I_p^{P_\frac12})}{w_{nc}^{p_\mp}(\mathcal{E},\omega,I_p^{P_\frac32})}}\\
\label{a2}
\frac{|a_c^{p_\pm\frac32}|}{|a_c^{p_\mp\frac32}|}&\approx&\sqrt{\frac{w_{nc}^{p_\pm}(\mathcal{E},\omega,I_p^{P_\frac32})}{w_{nc}^{p_\mp}(\mathcal{E},\omega,I_p^{P_\frac32})}}.
\end{eqnarray}
These ratios are easily evaluated using analytical expressions for energy-resolved ionization rates in Refs.\,\cite{PRA1,PRA2}.
Finally, the hole dynamics \eref{holedensityfinal} also depends on the geometrical shape of the valence atomic orbitals $4p_m$ ($m=0,\pm1$) of the krypton atom, i.e.
\begin{eqnarray}
\label{4pm}
4p_m&=&\frac{C_{41}}{2}\,r'e^{-\frac{r'}{4}}L_2^3\left(\frac{r'}{2}\right)Y_{1m}(\theta,\phi),
\end{eqnarray}
where $L_n^k(x)$ are Laguerre polynomials, $Y_{lm}(\theta,\phi)$ are spherical harmonics, $r'=Z_\mathrm{eff}\,r/a_0$ is the reduced radius, $\theta$, $\phi$ are the angles, and $C_{nl}$ is the normalization constant
\cite{Tipler,ringcurrent}.

\begin{figure}[ht]
\includegraphics[width=0.75\textwidth]{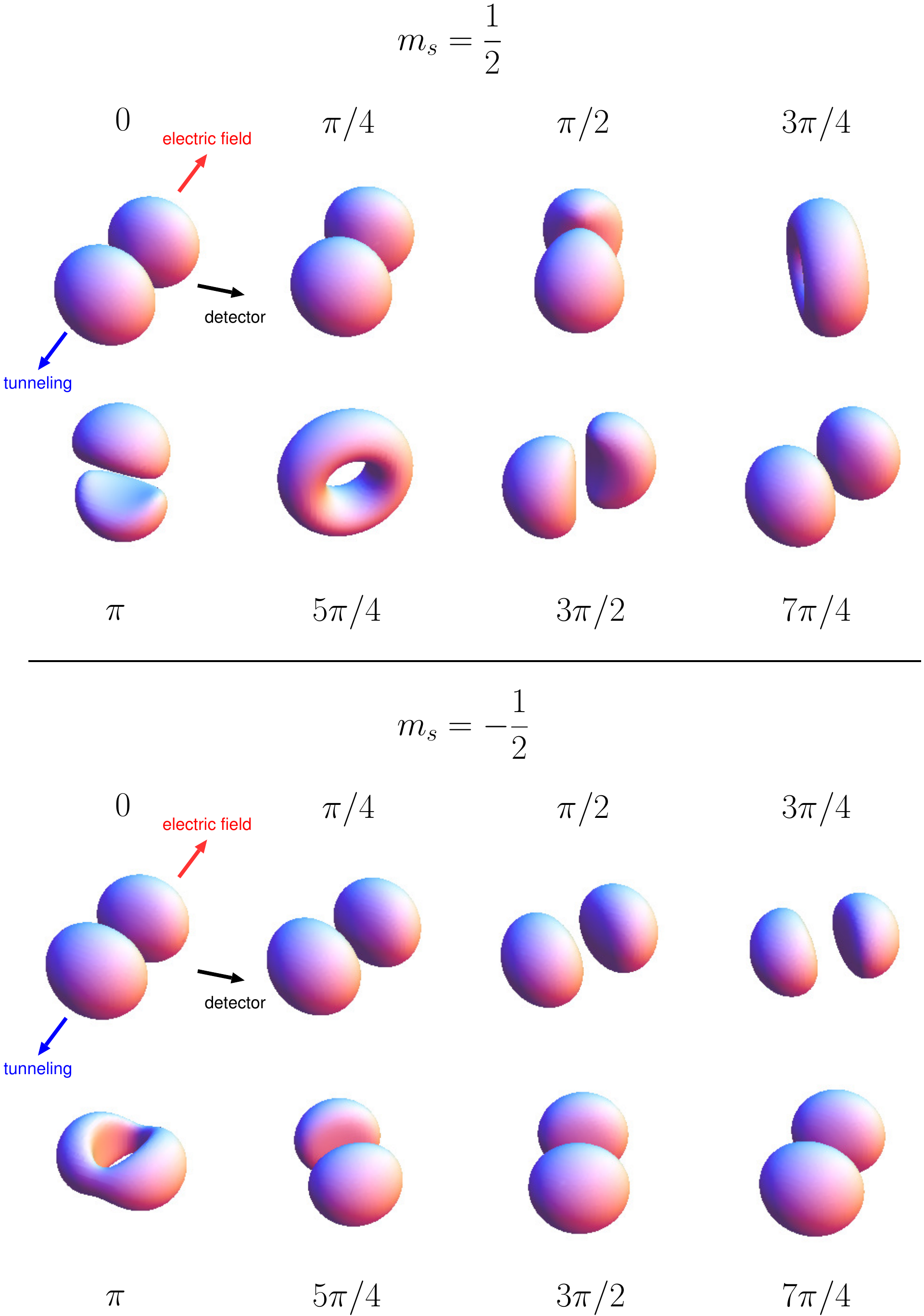}
\caption{}
\label{fig1}
\end{figure}

\begin{figure}[ht]
\includegraphics[width=0.75\textwidth]{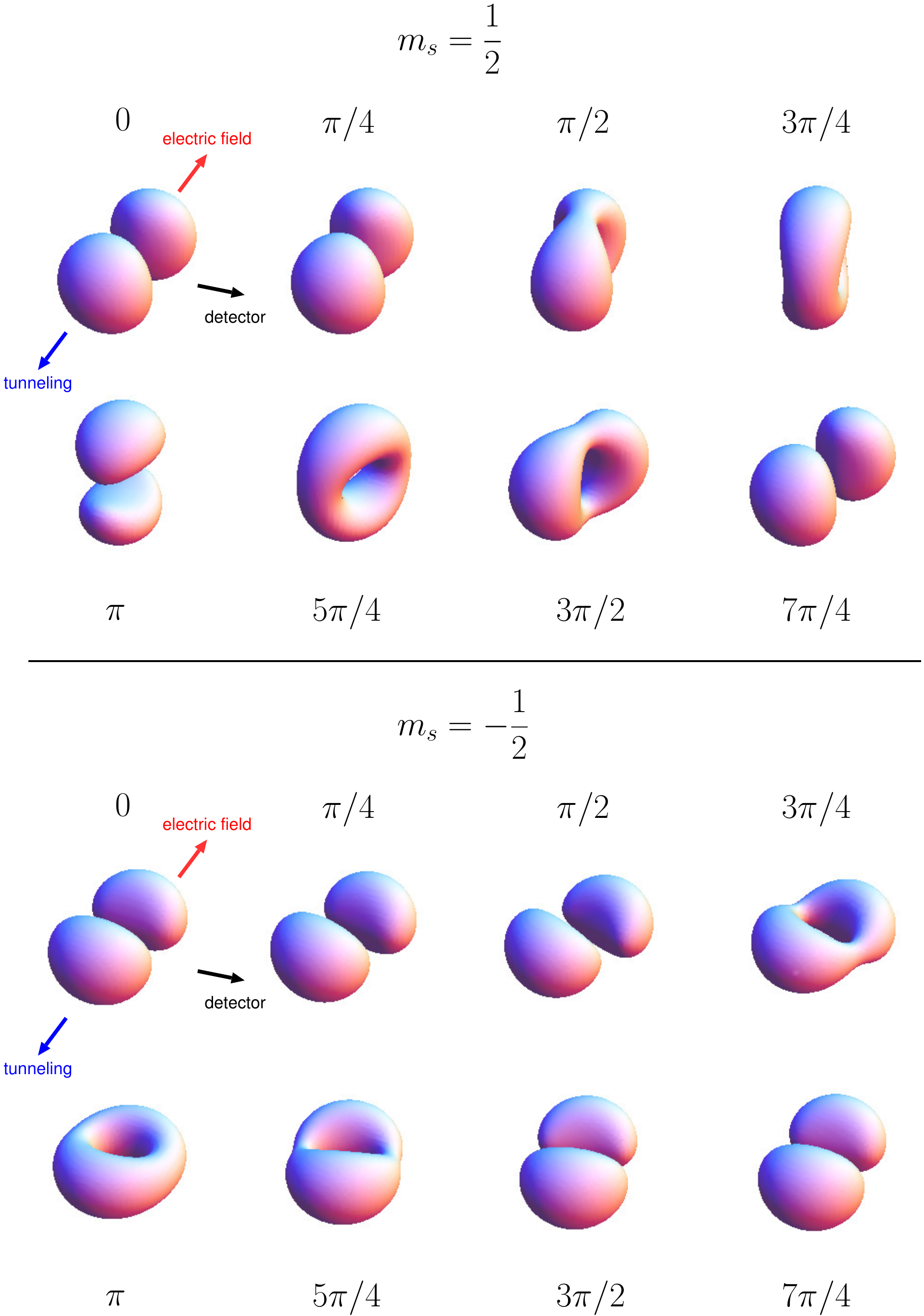}
\caption{}
\label{fig2}
\end{figure}

The time-dependent (phase-dependent) hole densities \eref{holedensityfinal} for both spins of the photoelectron $m_s=\pm\frac12$
are shown in \fref{fig1} for $\mathcal{E}=0.05$\,a.u., $\omega=0.057$\,a.u.\ (800\,nm), $E_\mathrm{kin}=0.70$\,a.u.\
and in \fref{fig2} for $\mathcal{E}=0.02$\,a.u., $\omega=0.0285$\,a.u.\ (1600\,nm), $E_\mathrm{kin}=0.45$\,a.u.
\Fref{fig1} and \fref{fig2}  illustrate  remarkable sensitivity of  the hole dynamics  to the laser parameters, the final kinetic energy and the spin of the photoelectron.

Snapshots shown in \fref{fig1} reveal complicated reshaping of hole density as it rotates in counter-clockwise direction for spin-up electron and clock-wise direction for spin-down electron.
\Fref{fig1} also illustrates an  initial alignment of the hole: the hole is aligned along the instantaneous direction of the laser field.

Snapshots shown in \fref{fig2} correspond to swinging motion of the hole, which starts in clock-wise direction for the spin-up electron and counter-clock-wise direction for spin-down electron.

Throughout this work we have focussed on the short-range part of the electron-core interaction.
Long-range effects in strong-field ionization in circularly polarized laser fields are described in  Ref.\,\cite{jivesh,archive}.
Rigorous analysis of the impact of these effects on the dynamics of the hole is outside the scope of the present work and is a subject of a separate detailed study.
However, the results of Ref.\,\cite{jivesh} allow us to estimate  the impact of the long-range effects on our results for the hole correlated to the electron
 with the "optimal" momentum, corresponding to the peak of the photoelectron energy distribution in case of many-cycle pulses.
Note that in short pulses, the effects of the pulse envelope significantly alter the momentum corresponding to the peak of the photoelectron distribution shifting it away from the 'optimal' momentum and making the analysis based on the optimal momentum irrelevant.
Here we do not consider short pulses and envelope effects. In this case, the hole dynamics is  determined by (i) the ratio of the ionization rates for electrons co-rotating and counter-rotating with respect to the laser field, and (ii) ratio of ionization rates of the electrons with the same sense of rotation to different final states of the ion, separated by $=0.67$\,eV.
The first ratio is weakly affected by the non-adiabatic Coulomb effects for long wavelengths of the driving circularly polarized laser fields, in particular 800\,nm ($\omega=0.057$\,a.u.) and 1600\,nm ($\omega=0.0285$\,a.u.), that we have used here (see figure 5 of Ref.\,\cite{jivesh}). The contribution of the short-range effects to the second ratio is dominant.
Thus, we do not expect significant Coulomb corrections to the hole dynamics correlated to the electrons corresponding to the  peak of the photoelectron energy distribution.
 However, we expect that the initial alignment of the hole will be affected by the non-adiabatic Coulomb effects, leading to (i) the angular off-set of about $10$ degrees (see figure 3 of Ref.\,\cite{jivesh}) for 800\,nm laser field due to the Coulomb corrections to the ionization time and (ii) an additional "geometrical" phase off-set associated with the Coulomb correction to the  ionization phase 
 (compare equation (97) for short-range and equation (98) for long range potentials in  Ref.\,\cite{jivesh}). The geometrical phase shift depends on laser parameters and it is smaller for our parameters than the angular off-set due to corrections to ionization time. All these estimates are only applicable to the holes, correlated to the photoelectrons with 'optimal' momentum.

\section{Conclusions}
\label{vii}
We have  derived an analytical formula for the spin-resolved hole densities after nonadiabatic ionization of the krypton atom in circularly polarized laser fields.
First, our analysis indicates that non-adiabatic effects in ionization leading to rotational dynamics of the hole are significant even for Keldysh parameter $\gamma\sim0.7$.
Second, our analysis has revealed significant impact of electron-hole entanglement on the hole motion: even in the case of short-range interaction of the electron with the core, the electron dynamics,  spin and energy strongly affect the dynamics of the hole.
Third, in short range potentials  the initial alignment of the hole  can be directly linked to the absence of time delays in its formation.
For long-range potential the connection between the initial alignment of the hole and ionization time can be more complicated.
Forth, we have shown that strong field ionization can create spin currents in the ion. Importantly, spin currents created after ionization should survive even after integration over the spin of the ejected electron, opening the way to studying magnetic effects in simple systems, such as noble gas ions produced by the strong laser field. 
 Fifth, the theory presented in this work is applicable to other noble gas atoms Ne, Ar, Xe, and Rn. In equations \eref{a1} and \eref{a2} one should simply use the corresponding values of the ionization potentials for each atom and use the same formulas to calculate the  ionization rates; the wavefunction \eref{4pm} has to be replaced by the generalized one for $np_m$ orbitals.
Finally, the analytical theory allows one to consider  single ionization event even for long laser pulses. Experimental separation of single ionization event requires application of very short pulses.
Such pulses are available experimentally \cite{ursi1,ursi2,ursi3}, but they may introduce additional envelope-dependent effects \cite{archive}, which are not considered here.


\ack
We thank DFG (projects Sm 292/2-1, Sm 292/2-3) gratefully for the financial support.

\section*{Appendix}

Here, we prove the relation for the $m$-dependent relative phase \eref{eta2} of the ionization amplitudes $a_c^{p_m}(\mathbf{p},t,I_p^{P_J})$ and $a_c^{p_{m'}}(\mathbf{p},t,I_p^{P_{J'}})$
for $m,m'=\pm1$, i.e.
\begin{eqnarray}
\label{eta2app}
\Delta\eta_2=\Delta\eta^{p_mp_{m'}}(\mathbf{p},t_i)&=&(m-m')\theta_p.
\end{eqnarray}
Since the ionization amplitude after applying the saddle point method is proportional to
the $m$-dependent term, i.e.\ the Fourier transformation of the wave function at the complex ionization time $t=t_i$
(cf. equations (35), (36), (42), (58), (63), and (65) of Ref.\,\cite{PRA2}),
\begin{eqnarray}
\label{acprop}
a_c^{p_m}(\mathbf{p},t,I_p^{P_J})&\propto&\sqrt{\frac{(1-m)!}{(1+m)!}}\,P_1^m(\cos\theta_{vc}(t_i))e^{im\phi_{vc}(t_i)},
\end{eqnarray}
where $\theta_{vc}(t_i)$ and $\phi_{vc}(t_i)$ are the spherical angles of the photoelectron velocity at $t=t_i$.
Since the first term $\sqrt{\frac{(1-m)!}{(1+m)!}}$ is purely real and positive, it does not contribute to the relative phase.
The associated Legendre polynomials $P_1^m(x)$ for $m=\pm1$ are related via
\begin{eqnarray}
\label{Leg}
P_1^{-1}(x)=-\frac12P_1^1(x)=\frac12P_1^1(x)e^{i\pi},
\end{eqnarray}
thus the corresponding relative phase between $m=-1$ and $m=1$ is $\pi$.
Since the so-called tunneling angle $\phi_{vc}(t_i)$ is complex, we have to rewrite the third term of equation \eref{acprop} for $m=\pm1$ as
\begin{eqnarray}
e^{im\phi_{vc}(t_i)}&=&\cos\phi_{vc}(t_i)+i\,\mathrm{sgn}(m)\sin\phi_{vc}(t_i)
=\frac{v_{xc}(t_i)+i\,\mathrm{sgn}(m) v_{yc}(t_i)}{v_{\rho c}(t_i)},
\end{eqnarray}
where $v_{xc}(t_i)=v_{\rho c}(t_i)\cos\phi_{vc}(t_i)$ and $v_{yc}(t_i)=v_{\rho c}(t_i)\sin\phi_{vc}(t_i)$.
With $\mathbf{v}_c(t)=\mathbf{p}+\mathbf{A}_c(t)$ and $\mathbf{A}_c(t)=-A_0(\sin(\omega t)\,\mathbf{e}_x-\mathrm{sgn}(c)\cos(\omega t)\,\mathbf{e}_y)$, we get
\begin{eqnarray}
e^{im\phi_{vc}(t_i)}
&=&\frac{p_x+i\,\mathrm{sgn}(m)\,p_{y}-A_0\sin(\omega t_i)+i\,\mathrm{sgn}(mc)\,A_0\cos(\omega t_i)}{v_{\rho c}(t_i)}.
\end{eqnarray}
With $p_x=p_\rho\cos\theta_p$, $p_y=p_\rho\sin\theta_p$, and $\omega t_i=\frac{\pi}{2}+\mathrm{sgn}(c)\,\theta_p+i\,\mathrm{arcosh}\,\chi$ \cite{PRA1,PRA2}, it yields
\begin{eqnarray}
\label{appeimphi}
e^{im\phi_{vc}(t_i)}&=&\frac{p_\rho -A_0\chi +\mathrm{sgn}(mc)\,A_0\sqrt{\chi^2-1}}{v_{\rho c}(t_i)}\,e^{im\theta_p}.
\end{eqnarray}
Finally, we have to show whether the sign of the numerator of equation \eref{appeimphi} is negative or positive, depending on $m=\pm1$.
Using $\mathbf{v}_c(t_i)^2=2p^2+A_0^2-2A_0\chi p_\rho$ and the saddle point equation $\mathbf{v}_c(t_i)^2/2+I_p=0$ \cite{PRA1,PRA2}, we obtain the expression for $\chi$, i.e.
\begin{eqnarray}
\label{chi}
\chi&=&\frac{A_0\alpha}{2p_\rho}\geq 1
\end{eqnarray}
with abbreviation
\begin{eqnarray}
\label{alpha}
\alpha&=&\left(\frac{p}{A_0}\right)^2+1+\gamma^2\geq 1,
\end{eqnarray}
where $\gamma=\sqrt{2I_p}/A_0>0$ is the Keldysh parameter.
Using equation \eref{chi}, equation \eref{appeimphi} is then rewritten as
\begin{eqnarray}
\label{appeimphi2}
e^{im\phi_{vc}(t_i)}&=&\frac{A_0\chi}{v_{\rho c}(t_i)}\left(\frac{\alpha}{2\chi^2} -1 +\mathrm{sgn}(mc)\,\sqrt{1-\frac{1}{\chi^2}}\right)\,e^{im\theta_p}.
\end{eqnarray}
There are no more than two zeros that are given by
\begin{eqnarray}
\chi_{0\pm}=\pm\frac{\alpha}{2\sqrt{\alpha-1}}.
\end{eqnarray}
We show using equations \eref{chi} and \eref{alpha} that the variable $\chi\geq1$ is always larger than $\chi_{0\pm}$, i.e. $\chi_{0\pm}<\chi$ due to $\sqrt{\alpha-1}=\sqrt{(p/A_0)^2+\gamma^2}>p/A_0$.
Thus, the sign of the term in the parenthesis in equation \eref{appeimphi2} can be determined by the sign of this term at $\chi\rightarrow\infty$.
Then, we can use Taylor expression $\sqrt{1-1/\chi^2}\approx 1-1/(2\chi^2)$. For $\chi\rightarrow\infty$, the term in equation \eref{appeimphi2} is positive for $mc=1$, i.e.
\begin{eqnarray}
\frac{\alpha}{2\chi^2} -1 +\sqrt{1-\frac{1}{\chi^2}}&\approx&\frac{\alpha-1}{2\chi^2}\geq0
\end{eqnarray}
and negative for $mc=-1$, i.e.
\begin{eqnarray}
\frac{\alpha}{2\chi^2} -1 -\sqrt{1-\frac{1}{\chi^2}}&\approx&\frac{\alpha+1}{2\chi^2} -2\approx -2\leq 0.
\end{eqnarray}
Hence, the corresponding relative phase between $m=-1$ and $m=1$ is $\pi$ that compensates the relative phase $\pi$ from the associated Legendre polynomials \eref{Leg}.
The remaining phase in equation \eref{appeimphi2} is $m\theta_p$, thus we have just proved the expression for the $m$-dependent relative phase \eref{eta2app}.

\newpage
\section*{FIGURE CAPTIONS}


\vspace{0.5cm}
\noindent
\textbf{Figure 1.}
Time-dependent (phase-dependent) hole densities (equation \eref{holedensityfinal})
for $m_s=\pm\frac12$, $\mathcal{E}=0.05$\,a.u., $\omega=0.057$\,a.u.\ (800\,nm), $E_\mathrm{kin}=0.70$\,a.u.
At the initial time $t=\mathrm{Re}\,t_i$, corresponding to the phase zero, the photoelectron that will reach the detector at the angle $\phi=\theta_p$ (black arrow) has to tunnel the barrier
at the angle $\phi=-\frac{\pi}{2}+\theta_p$ (blue arrow)
and therefore the initial hole is created (equation \eref{holedensityinitial}) when the vector of the right circularly polarized electric field is directed at the angle $\phi=\frac{\pi}{2}+\theta_p$ (red arrow).

\vspace{0.5cm}
\noindent
\textbf{Figure 2.}
Same as in Fig.\,2 but for $\mathcal{E}=0.02$\,a.u., $\omega=0.0285$\,a.u.\ (1600\,nm), $E_\mathrm{kin}=0.45$\,a.u.

\newpage
\section*{FIGURES}

\newpage


\begin{thebibliography}{100}
\bibitem{footnote}
The choice of the sign depends on the definition of spherical functions: the same definition must be used for derivation of ionization rates and for definition of spatial orbitals.
 We have used the following definition: $Y_l^m(\theta,\phi)=\left[\frac{(2l+1)}{4\pi}\frac{(l-|m|)!}{(l+|m|)!}\right]^{1/2}P_l^m(\cos(\theta))e^{im\phi}$.
\end{thebibliography}
\end{document}